\documentclass[12pt]{article}
\usepackage{graphicx}
\usepackage{subfigure}
\usepackage[dvips]{color}

\input epsf.tex

\newcommand{\beq}{\begin{equation}}
\newcommand{\eeq}{\end{equation}}
\newcommand{\be}{\begin{equation}}
\newcommand{\ee}{\end{equation}}
\newcommand{\bea}{\begin{eqnarray}}
\newcommand{\eea}{\end{eqnarray}}

\makeatletter

\@addtoreset{equation}{section}
\makeatother
\newcommand{\gs}{\mbox{$g_s$}}            
\newcommand{\ap}{\mbox{$\alpha^\prime$}}  
\newcommand{\ls}{\mbox{$l_s$}}            

\def\href#1#2{#2}

\def\p{\partial}

\begin{document}

\baselineskip=15.5pt
\pagestyle{plain}
\setcounter{page}{1}

\begin{titlepage}
\begin{flushleft}
       \hfill                       FIT HE - 11-02 \\
       \hfill                       KYUSHU-HET 131 \\
\end{flushleft}

\begin{center}
  {\huge Holographic Approach to Regge Trajectory   \\ 
   \vspace*{2mm}
and Rotating D5 brane \vspace*{2mm}
}
\end{center}

\begin{center}

\vspace*{2mm}
{\large Kazuo Ghoroku${}^{\dagger}$\footnote[1]{\tt gouroku@dontaku.fit.ac.jp},
Tomoki Taminato${}^{\ddagger}$\footnote[2]{\tt taminato@higgs.phys.kyushu-u.ac.jp},
\\
and ${}^{\P}$Fumihiko Toyoda\footnote[3]{\tt ftoyoda@fuk.kindai.ac.jp}
%
}\\

\vspace*{4mm}
{${}^{\dagger}$Fukuoka Institute of Technology, Wajiro, 
Higashi-ku} \\
{
Fukuoka 811-0295, Japan\\}
{
${}^{\ddagger}$Department of Physics, Kyushu University, Hakozaki,
Higashi-ku}\\
{
Fukuoka 812-8581, Japan\\}
{
${}^{\P}$School of Humanity-Oriented Science and
Engineering, Kinki University,\\ Iizuka 820-8555, Japan}

\vspace*{5mm}
\end{center}

\begin{center}
{\large Abstract}
\end{center}
We study the Regge trajectories of holographic mesons and baryons by considering
rotating strings and D5 brane, which is introduced as the baryon vertex. Our model is based on the type IIB superstring theory
with the background of asymptotic $AdS_5\times S^5$. This background is dual to a confining supersymmetric Yang-Mills theory
(SYM) with gauge condensate, $\langle F^2\rangle$, which determines the tension of the linear potential between the
quark and anti-quark. 
Then the slope of the meson trajectory ($\alpha'_{M}$) is given
by this condensate as $\alpha'_{M}=1/\sqrt{\pi \langle F^2\rangle}$ at large spin $J$. 
This relation is compatible with the other theoretical
results and experiments.
For the baryon, we show the importance of spinning baryon vertex to obtain a Regge
slope compatible with the one of $N$ and $\Delta$ series. 
In both cases, mesons and baryons,
the trajectories are shifted to
large mass side with the same slope for increasing current quark mass.

\noindent

\vfill
\begin{flushleft}

\end{flushleft}
\end{titlepage}
\newpage

\vspace{3cm}
\section{Introduction}
Among 
many observations of hadronic phenomena, the Regge behavior,
$ 
 J=\alpha_0+\alpha'_{\rm eff}M^2\, 
$, 
for the hadronic mass
$M$ and its spin $J$ has implied the existence of strings which connect 
constituent quarks. The string theory, 
however, has been  
established as the quantum theory of the gravity in ten dimensional
(10D) space-time. On the other hand,   
$SU(3)$ Yang-Mills theory has been established as
the basic one to form hadrons in 4D space-time.
So the two theories seem to be separated as different one.

After being uncovered these facts, new paradigm in the research of the strong 
coupling Yang-Mills theory has been opened by the conjecture known as gauge/gravity duality or AdS/CFT correspondence \cite{ads1,ads2,ads3}.
Many trials to understand strongly interacting gauge theories have been revived
within this paradigm. 
The approach is based on the 10D string theory, which describes gravitational theory.
Using the low energy effective action of this theory, many classical solutions have
been studied as the dual of the 4D Yang-Mills theories.
This approach is also called as the holography since the 4D gauge theory
is reflected in the higher dimensional gravity.

\vspace{.3cm}
Here we study the Regge behavior of hadrons from the gauge/gravity
duality. Two possible ways are considered in this approach.
One way is to examine 
the fluctuation
modes of probe brane embedded in the supergravity background dual to the
confining theory. {The approaches in this direction
have been performed from five dimensional
effective gravity \cite{TB,KKSS,ISS}. 
The string theory is however not yet solved in curved space-time, then it is still an open problem to introduce higher spin states
in a strict sense.}

Another way is to examine classical string-configurations in the supergravity background dual to the considering Yang-Mills theory. In this case, 
higher spin meson-states are introduced through the
rotation of the strings,
whose two end points are on the flavor brane introduced as a probe. 
While such states are not quantized, we can see the relation of mass and
spin for mesons in terms of these solutions at least for large spin. Here we study the Regge behavior through classical solutions
as performed in the references \cite{KMMW,ZSV,BCMZ,KZSV,PT,KV,Huang:2007fv}. 

In the ${\cal N}$=2 supersymmetric case with flavor brane(s) \cite{KMMW,PT,KV}, the 
background is given by $AdS_5$ and Regge-like behavior
is observed at small $J$ as
\beq\label{trajectory-1}
 J={1\over 2\pi m_q^2}
   \sqrt{\lambda\over \pi}M^2\, ,
\eeq
where $m_q$ and $\lambda$ denote the current quark mass and the 'tHooft coupling.
In this case, however, the Regge behavior is not observed at large $J$ because 
the potential between quark and anti-quark
is not linear at large distance. The 
behavior in the region of small $J$ is common to the theory dual to the 
background which
approaches to $AdS_5$ in the UV limit or near the boundary. 
Then this behavior corresponds to the behavior of the CFT. 
Here skipping this CFT behavior at small $J$, we concentrate on the Regge trajectory
given at large $J$ in the confining theory.

{In the background dual to the confining theory, 
we expect a linear Regge trajectory 
at large $J$. In the D4 branes background with flavor D6 probe brane model, such behavior has been observed \cite{BCMZ,KZSV,PT} at the limit of $m_q\to 0$,
\beq
  J=\alpha'\left({R_{D4}\over U_h}\right)^{3/2}M^2\, ,
\eeq
where {$\alpha'$, $R_{D4}$ and $U_h$ represent the string tension,
the radius of $S^4$ of bulk space-time
and the scale of compact $S^1$ of world volume of D4 brane respectively. }
It should be noticed that
main parameters of the bulk 10D background are remained in this formula.
For finite $m_q$, however, the trajectory deviates from the linear behavior.
This deformed behavior
has been discussed related to the heavy quarkonium \cite{KZSV}.  
In any case, as a common property of this kind of classical model,
$J$ arrives at zero for $M=0$, and then
the intercept of the trajectory is always
zero. In order to have a non-zero intercept, quantum corrections
would be necessary as pointed out in \cite{ISS,ZSV}.
Here we do not extend our analysis to such quantum version.
We study the Regge
trajectory by using only the classical solutions.}

\vspace{.5cm}
In our approach,
the 10D background configuration, which is dual to a confining 
supersymmetric Yang-Mills theory (SYM), 
is set as the solution of type IIB supergravity with five form field flux, 
dilaton and axion \cite{KS2,LT}. In this model, the tension $\tau_M$ of the linear potential
between the quark and anti-quark is given as $\tau_M=\sqrt{\langle F^2\rangle}/2$ by the gauge condensate $\langle F^2\rangle$  \cite{GY}.
The mesons and baryons are constructed by embedding probe D7 brane (flavor brane),
fundamental strings (F-strings) as quarks and D5 brane
as baryon vertex. The D7 brane is introduced to incorporate flavor quarks. 
The F-strings could end on this D7 brane.
These are all introduced as probes, so they do not change the
background configuration. That is, the vacuum of the gauge theory is not altered
by these branes.

The meson with higher spin is given by a rotating F-string whose both ends on the 
D7 brane. Its stable configuration is obtained by solving the equation of motion
for the Nambu-Goto action of the F-string in the bulk background.
In this case, appropriate boundary conditions for strings on the D7 brane are imposed. 
Since the mass and the spin of the meson 
are not quantized as mentioned above,
then they are continuous. At large $J$, as expected,
we find a linear trajectory,
\beq
  J={1\over \pi\sqrt{\langle F^2\rangle}}M^2\, ,
\eeq
with the slope
parameter $\alpha'_{M}=1/(2\pi\tau_{M})=1/{\pi\sqrt{\langle F^2\rangle}}$. 
{We should notice that the above formula is expressed only by 
the quantity of the gauge theory side without any parameter of the bulk side.
}
This
result is consistent with the picture obtained from naive quark model. 
That is, the rotating quark and anti-quark are connected by a string with the
tension of $\tau_M$. 
On the other hand,
at small $J$, we could
find the behavior found in the case of the $AdS_5$ background
mentioned above. The behavior in this region depends on the 
current quark mass $m_q$, 
which is determined by the profile function of the embedded D7 brane.\footnote{This 
is given by the value of the profile function at UV limit of the fifth coordinate ($r$).
However, in the present supersymmetric case, the profile function is a constant.}
As mentioned above, this meson configuration gives $J=0$ at $M=0$, then the intercept
is zero. 

In the next, we extend this approach to the baryon trajectories, where the baryons
are constructed according to the model proposed before \cite{wit}-\cite{GI}.
To make baryons, we must introduce D5 brane as the vertex which attracts $N_c$ F-strings.
The F-strings connect the D5 and D7 branes, and then both F-strings and D5 brane
contribute to the spin and mass in the case of the baryon.
While the D5 vertex could have various configurations,
for the simplicity, we examine here two typical configurations 
given in \cite{GI}. One is
called as ``point vertex'', which does not extend in the real three dimensional space.
So it looks like a point in our space. Another one is called as ``split vertex'', which
extends in one direction in our three space like a string. From the two end point of this
vertex, F-strings are going out toward the D7 brane. 

In the case of the point vertex, spin comes from the rotating
F-strings only since the point vertex couldn't generate rotation. 
As a result, we find that the Regge slope of this baryon is given by 
$\alpha'_{B}=2\alpha'_{M}/(N_c(1+\beta))$ with a positive constant $\beta$.
In this case, Then it may be difficult
to realize the desirable slope parameter in this case for $N_c\geq 3$. 

On the other hand, for the case
of split vertex, the smaller the energy of the state is, the shorter the attached strings
become, then the configuration may be dominated by the vertex.
The vertex volume
 doesn't shrink to zero and remains with a finite size then with a finite
energy.  
This means that 
the lowest energy state is finite contrary to the mesons.
Another point to be noticed is that the split vertex rotates to generate the spin.
So we estimate the spin and
the mass of the baryon by the vertex only in this case. 
Then the Regge slope of this configuration is given by the tension of the
split vertex.
We find a reasonable slope parameter
for the case of $N_c=3$ and $4$. It is favorite to consider the case of
$N_c=3$, but a problem to be solved is remained in the 
case of $N_c=3$ as discussed below. Here we add the analysis for $N_c=4$ since
this case is free from such a problem. 


The outline of this paper is as follows.
In the next section, the bulk solution for our holographic model is given.
In the section 3, meson trajectories are studied by using Nambu-Goto action.
Introducing D5 brane, the baryon trajectories are studied in the section 4.
The summary and discussions are given in the final section.

\section{Holographic Model}

Here we give the ingredient of our holographic model for supersymmetric
Yang-Mills theory with quarks to form meson and baryons.

\subsection{Bulk Background}
We consider 10D IIB model retaining the dilaton
$\Phi$, axion $\chi$ and self-dual five form field strength $F_{(5)}$.
Under the Freund-Rubin
ansatz for $F_{(5)}$, 
$F_{\mu_1\cdots\mu_5}=-\sqrt{\Lambda}/2~\epsilon_{\mu_1\cdots\mu_5}$ ,
and for the 10d metric as $M_5\times S^5$ or
$ds^2=g_{\mu\nu}dx^{\mu}dx^{\nu}+g_{ij}dx^idx^j$, the solution given below
has been found \cite{KS2,LT}.
Where $(\mu,\nu)= 0\sim4$ and  $(i,j)=5\sim9$.
The five dimensional part ($M_5$) of the
solution is obtained by solving the following reduced 5D action,
\beq\label{10d-action}
 S={1\over 2\kappa^2}\int d^5x\sqrt{-g}\left(R+3\Lambda-
{1\over 2}(\partial \Phi)^2+{1\over 2}e^{2\Phi}(\partial \chi)^2
\right), \label{5d-action}
\eeq
which is written 
in the string frame. 

\vspace{.3cm}
The solution is obtained under the ansatz,
\beq
\chi=-e^{-\Phi}+\chi_0 \ ,
\label{super}
\eeq
which is necessary to obtain supersymmetric solutions. 
The solution is expressed as
\bea 
ds^2_{10}&=&G_{MN}dX^{M}dX^{N}\nonumber\\
&=&e^{\Phi/2}
\left\{
{r^2 \over R^2}A^2(r)\left(-dt^2+(dx^i)^2\right)+
\frac{R^2}{r^2} dr^2+R^2 d\Omega_5^2 \right\} \ . 
\label{finite-c-sol}
\eea 
Then, supersymmetric solution is obtained as
\beq
e^\Phi= 1+\frac{q}{r^4}\ , \quad A=1\, ,
\label{dilaton}
\eeq
where $M,~N=0\sim 9$ and
$R=\sqrt{\Lambda}/2=(4 \pi g_s N_c{\alpha'}^2)^{1/4}=({\lambda{\alpha'}^2})^{1/4}$ and
$\lambda=4 \pi g_s N_c$ denotes the 'tHooft coupling. The dilaton is set as $e^{\Phi}=1$ at 
$r\to \infty$, and the parameter $q$ corresponds to the vacuum expectation value (VEV) 
of gauge fields strength~\cite{LT}
of the dual theory. Then this solution is dual to the four dimensional 
$\cal{N}$=4 SYM theory with a constant gauge condensate. Due to this condensate, the supersymmetry is reduced to $\cal{N}$=2 and then the conformal
invariance is lost since the dilaton is non-trivial as given above. As a result, 
the theory is in the quark confinement phase since
we find a linear rising potential between quark and anti-quark
with the tension $\sqrt{q}/(2\pi\alpha'R^2)$ \cite{KS2,LT,GY}.

Furthermore, we can see that the space-time is regular at any point.
In the ultraviolet limit, $r\to\infty$, 
the dilaton part $e^{\phi}$ approaches to one and 
the metric (\ref{finite-c-sol}) is reduced to $AdS_5\times S^5$. 
On the other hand, the dilaton part $e^{\phi}$ diverges in the infrared
limit $r\to 0$, so that one may expect a singularity at $r=0$. However there is
no such a singular behavior. 
This is assured by rewriting the metric (\ref{finite-c-sol}) 
in terms of new coordinate $z$, where $z=R^2/r$. Then we obtain 
\beq\label{background-2}
ds^2_{10}=e^{\Phi/2}{R^2\over z^2}
\left(-dt^2+(dx^i)^2+dz^2+z^2 d\Omega_5^2 \right) \ . 
\eeq 
In the infrared limit $z\to\infty$, we have
\beq
  e^{\Phi/2}{R^2\over z^2}=R^2\sqrt{\frac{q}{R^8}+\frac{1}{z^4}}\sim \frac{\sqrt{q}}{R^2}\, . 
\eeq
Therefore we find 10D flat space time in this limit and no singular point
\cite{KS2,LT}.

\vspace{.3cm}{
\subsection{D7 brane embedding}
In this background a probe D7 brane is embedded to introduce $\cal{N}$=2 hypermultiplet,
which corresponds to
a string connecting the stacked D3 branes and the probe D7 brane. This is regarded 
as the quark hereafter. The world volume of the D7 brane is set by rewriting  
the extra six dimensional part of (\ref{finite-c-sol}) as
\bea\label{D7-metric}
  \frac{R^2}{r^2} dr^2+R^2 d\Omega_5^2 &=&{R^2\over r^2}\left(dr^2+r^2d\Omega_5^2\right)
  ={R^2\over r^2}\left(\sum_{i=4}^{9}d{X^i}^2\right)\, \nonumber \\ 
  &=&{R^2\over r^2}\left(d\eta^2+\eta^2d\Omega_3^2+\sum_{i=8}^{9}d{X^i}^2\right)\, .
\eea
Then the world volume coordinate of D7 brane is taken as the four dimensional
space-time $x^{\mu}$ and extra four dimensions, $(\eta,\Omega_3)$. Then the induced
metric on the D7 brane is expressed as
\beq\label{RAD7}
 ds^2_{8}=e^{\Phi/2}R^2\left\{{r^2 \over R^2}A^2(r)\left(-dt^2+(dx^i)^2\right)+
{1\over r^2}\left((1+w'(\eta)^2)d\eta^2+\eta^2d\Omega_3^2\right)\right\}\, .
\eeq
where $r^2=\eta^2+w(\eta)^2$ and $w'(\eta)=\partial_{\eta}w(\eta)$ 
Here the D7 brane is embedded under the ansatz,
\beq\label{D7-sol}
  (X^8)^2+(X^9)^2=w^2(\eta)\,  .
\eeq
The solution of $w(\eta)$ is obtained as $w=$constant in the background 
considered here \cite{GY}. While, for the simplicity, the solution is given
as $(X^8,X^9)=(w,0)$ in \cite{GY}, 
{we can choose it at any point of the circle $ (X^8)^2+(X^9)^2=w$ since
the background is symmetric in the $(X^8,X^9)$ plane. Within the probe approximation,
$N_f<< N_c$, $N_f$ number of D7 branes may be put on this circle. We need here
at least two D7 branes, $N_f=2$, to obtain the split vertex given in the section 4.}
Anyway, for this solution,
we can make a simple image for the baryon configuration as shown in the section 4.
Remembering the relation, $z=R^2/r$,
the distance between D7 and D3 branes is given by a constant
$r_{D7}=R^2/z_{D7}$, then the current quark mass is obtained as 
\beq
  m_q={r_{D7}\over 2\pi\alpha'}={\sqrt{\lambda}\over 2\pi z_{D7}}\, .
\eeq   
}
\vspace{.3cm}
\subsection{Rotating string}
Mesons are given by a open string which two end points are on the D7 brane. The baryons
are made by $N_c$ strings, whose one end is on the D7 brane and the other end is on
a vertex for every string. The vertex is given by a probe D5 brane which wraps
on $S^5$ of extra dimension\cite{wit} in this model. 
Here the rotating string is formulated as follows.

In the following analysis we change
the coordinate $r$ as $R^2/r=z$, then we rewrite the metric (\ref{finite-c-sol}) as follows
\beq\label{background3}
ds_{10}^2=e^{\Phi/2}{R^2 \over z^2}
\left\{A^2(z)\left(-dt^2+(dx^i)^2\right)+
   dz^2+z^2 d\Omega_5^2 \right\} \ .
\eeq 
Further, the metric for the string which rotates around
the $x_3$ axis is rewritten by cylindrical polar coordinates as, 
\beq\label{background-susy}
ds^2_{(5)}=e^{\Phi/2}{R^2 \over z^2}\left(
 A^2(z)\left(-dt^2+d{\rho}^2+{\rho}^2d{\tilde{\theta}}^2+dx_3^2\right)+dz^2\right) \ . 
\eeq 
Taking the string world sheet as ${(\tau,\sigma)=(t,z)}$ and 
the ansatz, $\rho=\rho(z)$ and $\tilde{\theta}=\omega t$, the induced metric is 
given as
\beq
 g_{\tau\tau}=e^{\Phi/2}{R^2\over z^2}A^2(z)\left(-1+\omega^2\rho^2\right)\, , \quad
 g_{\sigma\sigma}=e^{\Phi/2}{R^2\over z^2}\left(A^2(z){\rho'}^2+1\right)\, ,
\eeq
where prime denotes the derivative with respect to $z$.
Then we have 
\beq\label{string-ac0}
 S_{\rm string}=\int dt {\cal L}=
     -{1\over 2\pi\alpha'}\int dtdz 
  e^{\Phi/2}A^2(z){R^2\over z^2}
        \sqrt{\left(1-\omega^2\rho^2\right)\left({\rho'}^2+A^{-2}(z)\right)}\, .
\eeq
From this, the spin $J_s$ and the energy $E_s$ of this string are given as
\beq\label{spin}
   J_s={\partial{\cal L}\over \partial\omega}
   ={1\over 2\pi\alpha'}\int dz e^{\Phi/2}A^2(z){R^2\over z^2}
        \omega\rho^2\sqrt{{\rho'}^2+A^{-2}(z)\over 1-\omega^2\rho^2}\, ,
\eeq
\beq\label{energy}
   E_s=\omega{\partial{\cal L}\over \partial\omega}-{\cal L}
   ={1\over 2\pi\alpha'}\int dz e^{\Phi/2}A^2(z){R^2\over z^2}
        \sqrt{{\rho'}^2+A^{-2}(z)\over 1-\omega^2\rho^2}\, .
\eeq
These are estimated by giving appropriate solutions for the corresponding
strings in mesons and baryons as shown below.

{Here we emphasize that the D7 and fundamental strings introduced
here are all treated as probes, so they do not alter the bulk background
and their embedded configurations. We only impose consistent
boundary conditions at their connecting points in constructing the hadron
configurations.
}
As for the D5 brane, we explain it in the section 4, where the baryons
are discussed.

\section{Regge trajectory of Meson}

The meson with higher spin is constructed by rotating F-string whose
end points are on the D7 brane. 
The configuration of such a meson state is obtained by solving the
equation of motion for the Nambu-Goto action (\ref{string-ac0})
with an appropriate boundary condition
at the D7 brane. Using this solution,
we find the relation of the spin (\ref{spin}) and the mass (\ref{energy})
to obtain the Regge trajectory.

\subsection{Rough Estimation of Slope parameter for small $m_q$}

It is difficult to obtain analytic solution for the string. So
before giving the numerical estimations,
we give a rough estimation of the relation $J\propto E^2$ in the case of light quark mass. At the small quark
mass limit, $m_q\to 0$, the speed of the quark approaches to the light velocity,
$\omega \rho_b\to 1$, where $\rho_b$ denotes the value of $\rho$ at the 
end points of the string. They are on the D7 brane.
In this limit, the integrations of (\ref{spin}) and (\ref{energy}) are
approximated as follows.

At first, we rewrite (\ref{spin}) and (\ref{energy}) by changing the 
integration variable from $z$ to $\rho$ as
\beq\label{Energy}
 E_{M}=2\int_0^{\rho_b} d\rho M(\rho )\, , \quad 
 M(\rho)={1\over 2\pi\alpha'}e^{\Phi/2}A^2(z){R^2\over z^2}
        \sqrt{1+A^{-2}(z)(\partial_{\rho}z)^2\over 1-\omega^2\rho^2}\, .
\eeq
\beq
 J_{M}=I\omega\, , \quad I=2\int_0^{\rho_b} d\rho~\rho^2 M(\rho )\, .
\eeq
Here we can regard $M(\rho)$ and $I$ as the mass density and the inertial moment
of the rotating string respectively. Since $M(\rho)$ depends on $\rho$ through $z(\rho)$,
which is obtained as the solution of the equations of motion, it is hard
to perform the integration in (\ref{Energy}). In the present case, we can 
however estimate
it by a reasonable approximation, which is ensured by the numerical estimations
given later. 

$M(\rho)$ is written as
\bea
  M(\rho)&=&{\cal M}_{\rm sta}\sqrt{1\over 1-\omega^2\rho^2}\, ,\\
 {\cal M}_{\rm sta}&=&{1\over 2\pi\alpha'}e^{\Phi/2}A^2(z){R^2\over z^2}
        \sqrt{1+A^{-2}(z)(\partial_{\rho}z)^2}\, .
\eea
We notice that $M_{\rm sta}$ is the integrand in estimating the potential of
the static quark and the anti-quark. As we have previously shown, we find a linear
potential in our present model at large distance between the quark and the anti-quark.
This implies, for large $\rho_b$, that we obtain
the following result
\beq
  E_{\rm sta}=2\int_0^{\rho_b} d\rho {\cal M}_{\rm sta}=2\tau_{M}\rho_b\, , 
\eeq
where the tension, $\tau_{M}$, is given in our model as follows \cite{GY}
\beq\label{t-meson}
 \tau_M={\sqrt{q}\over 2\pi{\alpha'}R^2}\, .
\eeq

This result is understood as follows. For large $\rho_b$, 
the string shape approaches to the U shape
with a long bottom horizontal-line, which is the main part of the string
configuration. In the range of
this part, $R^2/z=r$ is small and almost constant. Then
in this region, we can set as
\beq
  {\cal M}_{\rm sta}
       \simeq {\sqrt{q}\over 2\pi{\alpha'}R^2}=\tau_{M}
\eeq
for the supersymmetric case (\ref{dilaton}), which is written also as, 
\beq\label{susy-sol-2}
e^\Phi= 1+{\tilde{q}}{z^4}\ , \quad\ \tilde{q}={q\over R^8}\, , \quad A=1\, .
\eeq 
Hereafter, we consider this solution for the simplicity.
In this approximation, the both side parts 
of U-shaped string configuration are
neglected. They corresponds to the effective quark and anti-quark masses.
So the approximation is applied to the case of small quark mass.

Here we notice the following relation \cite{LT}
\beq
 q=\pi^2\langle F^2\rangle \lambda\alpha'^4\, ,
\eeq
where $\lambda=4\pi g_s N_c$ and $\langle F^2\rangle$ denote the 'tHooft coupling 
and the VEV of the gauge field condensate. 
Then, remembering $R^4=\lambda\alpha'^2$, we find
\beq
 \tau_M={1\over 2}\sqrt{\langle F^2\rangle}
\eeq
{This implies that the tension does not depend on the other parameters,
$R$ and $\alpha'$, and then it is determined only by the mass scale,
$\langle F^2\rangle$,  of the confining gauge theory. We notice here another mass scale, $R$, which
determines the meson mass obtained from the D7 brane fluctuations \cite{BGN}. We 
also observe
the Regge behavior with this mass scale 
when we restrict the region to the small mass and to the finite quark mass case.
In this case, the trajectory
is given by (\ref{trajectory-1}) as mentioned in the Introduction. This behavior
is seen as a common properties of the meson spectra through the probe brane
fluctuations.
}

\vspace{.2cm}
Then ${\cal M}_{\rm sta}$ can be approximated by the QCD-string tension $\tau_{M}$
for large $J$ and small quark mass.
Under this approximation, $E_{M}$ and $J_{M}$ are estimated as follows,
\beq
  E_{M}=2\tau_{M}\int_0^{\rho_b} d\rho \sqrt{1\over 1-\omega^2\rho^2}\, , \quad 
  I=2\tau_M\int_0^{\rho_b} d\rho~{\rho^2 \over \sqrt{1-\omega^2\rho^2}}\, . 
\eeq
and using the approximation $\omega \rho_b\simeq 1$, we get
\beq\label{meson-trajectory}
  E_{M}=\pi\tau_{M}\rho_b\, , \quad J_{M}={\pi\over 2}\tau_{M}{\rho_b}^2\, .
\eeq
As a result, we obtain
\beq\label{meson-trajectory2}
  J_{M}={1\over 2\pi\tau_{M}}{E_{M}}^2\ .
\eeq
This implies that the slope of the meson trajectory is given by
\beq\label{meson-trajectory3}
 \alpha^{\rm Regge}_{M}={1\over 2\pi\tau_{M}}={1\over \pi\sqrt{\langle F^2\rangle}}\, .
\eeq
The result (\ref{meson-trajectory2}) is interesting since its form is parallel
to the one obtained
in the flat bulk-space analysis \cite{ZSV}, where the Regge slope is obtained as
\beq
  \alpha^{\rm Regge}_{M}={1\over 2\pi T_0}\, , \quad 
  T_0={1\over 2\pi\alpha'}\, .
\eeq
We find that our result is obtained by replacing the F-string tension $T_0$
by the QCD-string tension $\tau_{M}$ in the above formula. This point is
well understood from the fact that we find 10D Minkowski space-time in the 
IR limit of our model as shown above. However, 
we should notice that the scale of the coordinates are 
redefined as $(q^{1/4}/R) x^{\mu}\to x^{\mu}$.

\subsection{Quark mass effect}

In the next, we consider the case with a finite $m_q$, which can not be
negligible.
We rewrite the lagrangian (\ref{string-ac0}) according to the
formulation performed in \cite{KZSV} as follows,
\beq\label{string-ac0-1}
 {\cal L}=\int_{z_-}^{z_+} {\it L(z)}dz
    =\int_{z_-}^{z_l}{\it L(z)}dz+\int_{z_l}^{z_r}{\it
    L(z)}dz+\int_{z_r}^{z_+}{\it L(z)}dz\ ,
\eeq
where
\beq
     {\it L(z)}=-{1\over 2\pi\alpha'} P(z)
        \sqrt{\left(1-\omega^2\rho^2\right)\left({\rho'}^2+1\right)}\, ,
\eeq
and
\beq
        \quad P(z)=e^{\Phi/2}A^2(z){R^2\over z^2}\, .
\eeq
The coordinates $z_{\pm}$ denotes the two end points of the F-string at the D7
flavor brane. The region (i) $z_l<z<z_r$ represents the bottom range of the U-shaped
string configuration, which is considered in the above subsection. Then, the two
parts, (ii) $z_-\leq z \leq z_l$ and 
(iii) $z_r\leq z \leq z_+$, are corresponding the rotating quark
and anti-quark respectively. Since the configuration is symmetric, in the 
present case, then they are equal. And they are neglected in the above as very small
quantity. Here we estimate them.

We solve the string equations in terms of the above lagrangian,
(\ref{string-ac0-1}). The solution has the U-shape, so we approximate
as 
\beq
  |\rho'(z)|\gg 1\, , 
\eeq
for region (i)
\beq
  \rho'(z)\sim 0\, , 
\eeq
for region (ii) and (iii). Then the variation with respect to $\rho$ is
given as
$$
  \delta {\cal L}=-{1\over 2\pi\alpha'}\int_{z_l}^{z_r}dz\left\{ -\partial_z\left(
      P(z)\sqrt{1-\omega^2\rho^2}{\rho'\over \sqrt{1+{\rho'}^2}}\right)-
      P(z)\sqrt{1+{\rho'}^2}{\omega^2\rho\over \sqrt{1-\omega^2\rho^2}}
  \right\}\delta\rho $$
$$
 -{1\over 2\pi\alpha'}\left[P(z)\sqrt{1-\omega^2\rho^2}{\rho'\over \sqrt{1+{\rho'}^2}}
 \delta\rho\right]_{z_l}^{z_r}
$$
\beq\label{string-ac0-2}
   -{1\over 2\pi\alpha'}\left\{\int_{z_-}^{z_l}dz~P(z){-\omega^2\rho\over \sqrt{1-\omega^2\rho^2}}\delta\rho+\int_{z_r}^{z_+}dz~P(z){-\omega^2\rho\over \sqrt{1-\omega^2\rho^2}}\delta\rho
   \right\}\ .
\eeq
The second boundary terms come from the partial integration of the variation
for the first term of (\ref{string-ac0-1}). This must be cancelled out with
the terms in the third parenthesis of (\ref{string-ac0-2}) since the first term
vanishes due to the equation of motion. Noticing the symmetric configuration,
we find the next condition,
\beq
 1-\omega^2\rho^2_b=2\pi\alpha'{\omega^2\rho_b\over P(z*)}m_q\, ,
\eeq
\beq\label{quark-mass}
   m_q=-{1\over 2\pi\alpha'}\int_{z_r}^{z_+}dz~P(z)\ ,
\eeq
where $m_q$ denotes quark mass. Here, further
we set as $z_l=z_r=z*$ and $\rho(z*)=\rho_b$. The negative sign of the above
(\ref{quark-mass}) comes from $\rho'(z_r)\sim -\infty$. We notice that the position
of the D7 brane is given by $r_f=R^2/z_{\pm}$. Since $z_+<z_r$, we have
positive $m_q$. The same relation between $\rho_b$ and $m_q$ is obtained in 
from the other side condition. This relation is consistent with the fact that
the speed of the massless quark is one, namely $\omega\rho_b=1$ for $m_q=0$.

\vspace{.3cm}
Then, under the condition of small $m_q$ but non-zero, we estimate $J_M$ and
$E_M$ according to (\ref{spin}) and (\ref{energy}) by separating the region of
$z$ as above. Using the condition given above at the separating point, we obtain
\beq
   E_M=\pi\tau_M(\rho_b+\gamma\rho_b^{1/2})\, ,
\eeq
\beq
   J_M={1\over 2}\pi\tau_M\left(\rho_b^2+{2\gamma}\rho_b^{3/2}\right)\, ,
\eeq
where
\beq
 \gamma={2\over \pi}\sqrt{{2\pi\alpha' m_q\over P(z^*)}}\, .
\eeq
Then $J_M$ is expressed by $E_M$ by eliminating $\rho_b$. It is further
expanded as a series of $\gamma$ as
\beq
  J_M={E_M^2\over 2\pi\tau_M}-{\gamma^2\over 2\pi^2}E_M+O(\gamma^3)\, .
\eeq
The second term is negative and proportional to $m_q$.
Then we would find a shift of the trajectory to the right in the $J-E^2$ plane
as the quark mass effect. We should notice, however, that the slope given
at large $E_M$ is not changed by $m_q$.

\subsection{Numerical estimations}

\vspace{.3cm}
\noindent{\bf Equations of motion for SUSY background}

\vspace{.3cm}
The values of these quantities are obtained by introducing the D7 brane
to consider quarks with the finite mass. Here, we consider the supersymmetric case,
(\ref{dilaton}) or (\ref{susy-sol-2}).
In order to solve the string equation, it is convenient to use the reparametrization
invariant formalism. The solutions $\rho(z)$ are multi-valued for $z$, however the 
configuration of one solution is given by a continuous curve. So the solution can be
expressed by one parameter, $s$ as given here.
The Lagrangian is written by using the parameter $s$ as
\beq\label{effL}
{\cal L}=-{1\over 2\pi\alpha'}\int ds {\tilde L}
     =-{1\over 2\pi\alpha'}\int_{s_l}^{s_r} ds 
  e^{\Phi/2}A^2(z){R^2\over z^2}
        \sqrt{\left(1-\omega^2\rho^2\right)\left(\dot{\rho}^2+\dot{z}^2A^{-2}(z)\right)}\, .
\eeq
where dot denotes the derivative with respect to $s$. $s_l$ and $s_r$ are defined
as
\beq\label{bc-1}
  z(s_l)=z(s_r)=z_{D7}\, , \quad \dot{\rho}(s_l)=\dot{\rho}(s_r)=0\, .
\eeq
The first two equations means that the end points of the string are on the 
D7 brane, then the second equations are coming from the Dirichlet condition
$\rho'(z_{D7})=0$ for the string end points on the D7 brane. We impose these
boundary conditions in solving $z(s)$ and $\rho(s)$ by using (\ref{effL}).

The equations to be solved are obtained as follows. Introducing the canonical
momentum as,
\beq
  p_{\rho}={\partial {\tilde L}\over \partial\dot{\rho}}\, , \quad
  p_{z}={\partial {\tilde L}\over \partial\dot{z}}\, ,
\eeq
we have the Hamiltonian
\beq
  H=2{\tilde{H}\over \Delta}\, , \quad 
    \Delta={F\over \sqrt{\dot{\rho}^2+\dot{z}^2A^{-2}(z)}}\, ,
\eeq
\beq
  F= e^{\Phi/2}A^2(z){R^2\over z^2}
        \sqrt{1-\omega^2\rho^2}\, ,
\eeq
\beq
  \tilde{H}={1\over 2}\left(p_{\rho}^2+p_{z}^2A^2(z)-F^2 \right)\, .
\eeq

Then the Hamilton equations are obtained from $\tilde{H}$ instead of $H$
for the simplicity,
\beq\label{seq-1}
  \dot{\rho}=p_{\rho}\, , \quad \dot{z}=p_{z}A^2(z)\, , \quad 
\eeq
\beq\label{seq-2}
  \dot{p}_{\rho}=-\omega^2{\rho}Q(z)\, , \quad 
  \dot{p}_{z}=-p_z^2A(z){\partial A(z)\over \partial z}+
{1\over 2}\left(1-\omega^2\rho^2\right){\partial Q(z)\over \partial z}\, , \quad 
\eeq
and
\beq
  Q=e^{\Phi}A^4(z){R^4\over z^4}\, .
\eeq

In solving the above simultaneous equations (\ref{seq-1}) and (\ref{seq-2}), 
the boundary conditions (\ref{bc-1}) are imposed as mentioned above. And we
need the values of the parameters used in the equations to get numerical results.

\vspace{.5cm}
\noindent{\bf About the value of $\lambda=R^4/{\alpha'}^2$}

We expect holographic approaches to be useful at large $\lambda=R^4/{\alpha'}^2$ and 
large $N_c$. Although, in our real world, $N_c=3$ and this is not large enough,
we dare to compare our results with the experiments obtained up to now. We know
the $\rho$ meson trajectory, which is approximated by $J=0.53+0.88 E^2$. So, 
from (\ref{meson-trajectory3}), we can set as
\beq\label{meson-trajectory4}
 \alpha^{\rm Regge}_{M}={1\over \pi\sqrt{\langle F^2\rangle}}=0.88~({\rm GeV}^{-2})\, .
\eeq
This implies
\beq
  \sqrt{{\langle F^2\rangle}}=0.36~({\rm GeV}^{2})\, .
\eeq
This is our prediction for the gauge condensate. As for the value of this
quantity, many attempts to estimate it
have been done in 4D gauge theory side. We compare our result with
the one given in \cite{EGM},
\beq
  {\langle {g_{\rm YM}^2N_c\over 4\pi^2}F^2\rangle}=0.14~({\rm GeV}^4)\, ,
\eeq
according to the lattice simulation. Here the value is obtained for quenched
approximation and for $N_c=3$. Our calculation is also quenched for the quarks,
then they can be compared. Noticing $g_{\rm YM}^2N_c=4\pi g_s N_c=\lambda$,
we find 
\beq\label{lambda}
 \lambda=42\, .
\eeq
Then $\lambda$ is not small, thus it may be reasonable to compare our result with 
the quantities in our real world.

\vspace{.3cm}
{In applying the above result (\ref{lambda}), we should notice that
our model is based on the self-dual gauge field
configurations which satisfies $FF=F\tilde{F}$. Then we should 
study also the value of
$F\tilde{F}$ to check the validity of using the result given
in the lattice theory, where the supersymmetry is however lost. 
As in the case of $FF$, we have searched for simulations of $F\tilde{F}$ in lattice
theories, but we could not find any definite result for
this quantity. The lattice formulation might be unsuitable to study topological quantities like
$F\tilde{F}$.

Even if the self-dual relation for the gauge fields in the background 
were largely broken in the real QCD, it would be still meaningful to show our analysis
since we expect that our main results would survive
as being approximately true in the real QCD. Although this point is open, our
results are not so sensitive to
the value of $\lambda$ given above as shown below.
}

\vspace{.3cm}
\noindent{\bf Numerical results}

\begin{figure}[htbp]
\vspace{.3cm}
\begin{center}
\includegraphics[width=12cm]{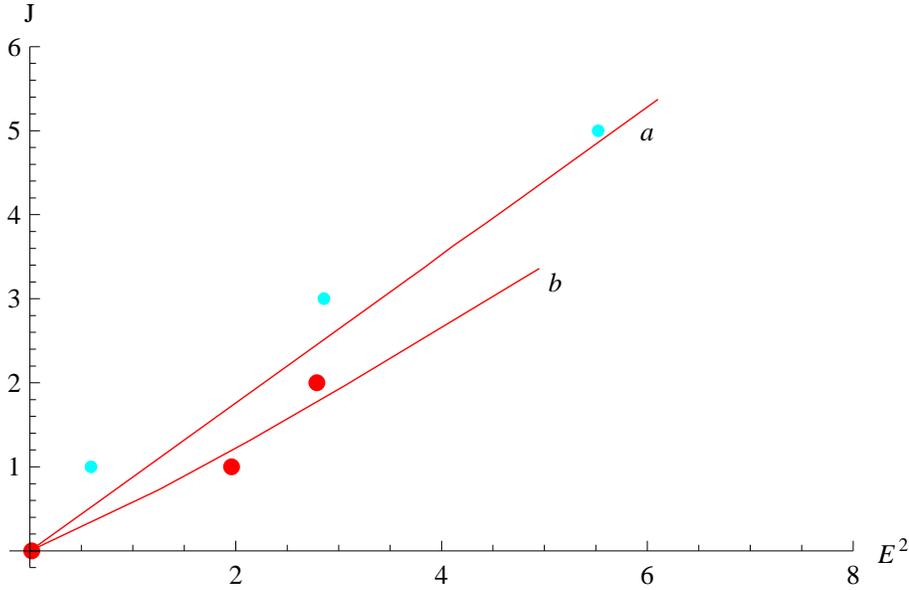}
\caption{{\small The numerical result of $J-E^2$ relations for $\tilde{q}=1/(.88^2\times 42)=0.0306$ and $R^2/\alpha'=6.48$. 
The curves (a) and (b) denote the results for $1/z_{\rm D7}=0.005$,
and $0.50$ respectively. Here the curves are almost the same for $20\leq\lambda\leq 42$.
And the small (large) dots denote the $\rho$ ($\pi$) meson series.}}
\label{rho-trajectory}
\end{center}
\end{figure}
\vspace{.3cm}
In obtaining the numerical values of $J$ and $E$, we must fix 
the position of the D7 brane or the quark mass $\lambda^{1/2}/(2\pi z_{\rm D7})=m_q$, and 
other parameters,
$\tilde{q}$ ( or $q$) and $R^2/\alpha'$ (or $\lambda$). The latter two
are fixed by neglecting the quark mass and using
our formula, experimental $\rho$ meson trajectory and 
a result of lattice simulation. The results are
given in the previous subsection (3.2).


In Fig.\ref{rho-trajectory}, two numerical results are shown 
for $\tilde{q}=0.0306$ and $R^2/\alpha'=6.48$ with the mass spectrum of 
$\rho$ and $\pi$ meson series. 
The two curves, (a) and (b) are for $1/z_{\rm D7}=0.05$ and $0.5$,
respectively, where $1/z_{\rm D7}$ denotes the position of the D7 brane
and it is equivalent to the quark mass $m_q$. 
For very small $m_q$ the trajectory is linear in almost all region
and is consistent with (\ref{meson-trajectory3}). Further
we find the expected shift to the right of the trajectory for the large $m_q$ case,  $1/z_{\rm D7}=0.5$.
As for the quark mass dependence, we can see that the heavier the quark mass becomes,
the larger the energy is for the same $J$. This is a reasonable result.
The value of $\lambda$ is set as 42 in the figure, we notice however that
the results are the same for wide range of $\lambda$, $20<\lambda<42$. 

{We notice one more important point that the slopes of the two curves are different at small mass region.
While, for any quark mass, we could find tha same slope given by (\ref{meson-trajectory3}) at large meson mass
as discussed above, we can see the difference of the slope, which depends on the quark mass $m_q$, in the
region of small meson mass.  This behavior in small meson mass region is consistent with the trajectory
given by (\ref{trajectory-1}). In this sense, then we could see two regions where the mass scale of the string
tension is different.}

However we should notice that both of the curves
approach to the origin for $J\to 0$. This implies that the intercept of the trajectory
gives zero in the present case. 
This is understood as follows. In our calculation,
the classical solutions of the Nambu-Goto action are used, and they are U-shaped
and give positive $E$. When $E$ vanishes, the string configuration also disappears
since the calculation is classical. 
Then, $J$ vanishes simultaneously for $E=0$. This implies that the intercept
should be zero in this calculation.
One way to get finite intercept,
in this approach, is to
add the quantum effect for the string configurations obtained here. 
In this case, we could expect to 
get a positive intercept as given for the closed string
case in \cite{ZSV}. We, however, do not consider the quantum 
corrections in this paper, and
we concentrate on the slope parameter of the Regge trajectory at large 
$E$ and $J$, where the
classical approximation would be good.

\section{Baryon Regge trajectory and rotating vertex}

{The baryon is constructed from the vertex and $N_c$ fundamental strings.
The vertex is given by the D5 brane which wraps 
${S}^{5}$ of the metric (\ref{finite-c-sol}) (or (\ref{background-2})) given above and 
couples to the 5 form fluxes of the bulk through the $U(1)$ gauge fields in it as shown below.
Due to this dissolved fluxes, the baryon vertex could have non-trivial configurations.

In order to show explicit configurations of the baryon vertex considered here,
we parametrize the $S^5$ by the five angle variables, $\left\{\theta,\theta_2,
\theta_3,\theta_4,\theta_5\right\}$. Then the coordinate $X^i (i=4\sim 9)$
of (\ref{D7-metric}) are written as,
\bea
  X^9&=&r\cos\theta\, , \nonumber \\
  X^8&=&r\sin\theta\cos\theta_2\, , \nonumber \\
  X^7&=&r\sin\theta\sin\theta_2\cos\theta_3\, , \nonumber \\
  X^6&=&r\sin\theta\sin\theta_2\sin\theta_3\cos\theta_4\, , \nonumber \\
  X^5&=&r\sin\theta\sin\theta_2\sin\theta_3\sin\theta_4\cos\theta_5\, , \nonumber \\
  X^4&=&r\sin\theta\sin\theta_2\sin\theta_3\sin\theta_4\sin\theta_5\, , 
\eea
and we have 
$\eta^2=\sum_{i=4}^{7}{X^i}^2=r^2\sin^2\theta\sin^2\theta_2$.

In general, we can consider many kinds of configurations for the
D5 brane. They will be shown by introducing the external coordinates 
on the D5 brane as, $r(\theta_i)$
and $X^i(\theta_j), i=1\sim 3$, but it would be very complicated to solve
the equations of motion to minimize the action for these general functions. 
Then, we restricted our analysis to
a simple case
with $r(\theta)$ and $(X^1(\theta))^2+(X^2(\theta))^2\equiv \rho^2(\theta)$. In this sense,
our configurations adopted here are not special but simple. For more complicated
configurations, they are remained as future works.

The two 
typical configurations of baryons considered here
are shown in the Figs.\ref{d5-d7-point} and \ref{d5-d7-split} in the three
dimensional space of
$\left\{X^9,X^8, \eta\right\}$, where 
the distance from the origin
is equivalent to the coordinate $r$, which is denoted as
$r = \sqrt{(X^9)^2+(X^8)^2+ {\eta}^2}$.  

As mentioned above,
the D5 brane is embedded to minimize its action with the coordinates
$\left\{r(\theta), \rho(\theta)\right\}$, where $\rho^2=(X^1)^2+(X^2)^2$.
Unfortunately, in the Figs.\ref{d5-d7-point} and \ref{d5-d7-split},
we can not see the coordinate $\rho$ and then the extension
of $\rho(\theta)$ for the split vertex. The point and split vertices are 
discriminated here by the number of the cusp, namely one for the point and
two for the split as shown in the Figs. \ref{d5-d7-point} and \ref{d5-d7-split}.

As for the D7 brane, its configuration
is expressed by the solution given in (\ref{D7-sol}) in the form of a tube of $w=$constant.
{We should notice here about the number of the D7 branes.
In the Fig. \ref{d5-d7-point}, the point vertex is shown with one cusp
of D5 brane. In this case, we need one D7 brane at $(X^8,X^9)=(0,w)$
as the end point of strings
as shown in the figure. On the other hand, we need at least two flavor D7 branes
at $(X^8,X^9)=(0,\pm w)$ for the split vertex baryon as shown in 
the Fig. \ref{d5-d7-split}. This implies that the number of the flavor quarks
might be related to the 
configuration of the baryon vertex,
especially to the number of the cusps on the D5 brane. This point should be examined
more by considering more complicated configurations in the future.}

\begin{figure}[htbp]
\vspace{.3cm}
\begin{center}
\includegraphics[width=7cm]{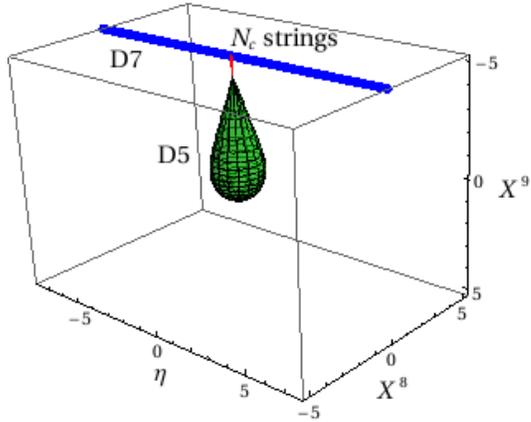}
\caption{{\small The typical configuration of point vertex(D5 brane) and flavor brane(D7 brane) in the space ($X^8$, $X^9$, $\eta$). The point vertex has one cusp 
at $\theta=\pi$, and then all $N_c$ strings should attach to this cusp.}}
\label{d5-d7-point}
\end{center}
\end{figure}
\begin{figure}[htbp]
\vspace{.3cm}
\begin{center}
\includegraphics[width=14cm]{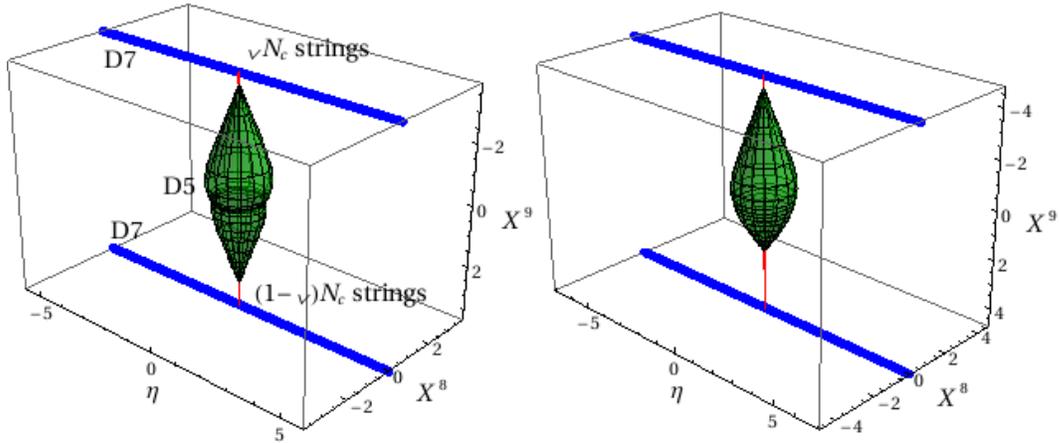}
\caption{{\small The typical configurations of split vertex and two flaver branes on ($X^8$, $X^9$, $\eta$). The split vertex has two cusp at $\theta=0,\ \pi$. $\nu N_c$ strings should attach to the cusp at $\theta=0$, and $N_c(1-\nu)$ strings should attach the other cusp. Two shapes of the vertex are shown. They 
depend on the boundary conditions of embedding equations, the left one with a spindle-shape and the right with the peanuts form in the figure.}}
\label{d5-d7-split}
\end{center}
\end{figure}

As mentioned above, both D branes are embedded as probes. Then we neglect
the back reactions. However, we should notice that there are cusps in D5 brane
due to the dissolved bulk flux as $U(1)$ fields in it. On the other hand,
the fundamental strings, which terminate on these cusps, are dissolved
in the D5 brane \cite{wit,groguri} as the $U(1)$ flux. Then there is a smooth
flow of the strings from D5 brane vertex.

On the other hand, the embedded D7 brane
has no cusp, and then strings end on this brane with the Diriclet conditions. This
situation is the same with the case of mesons, which are given as strings whose
two end-points are on the D7 branes with the same boundary condition. For the case
of the baryon, however, the situation would be slightly different since
the baryon number is not zero due to the dissolved strings. 
As a result, the $U(1)$ gauge field in the D7 brane would be a non-trivial
configuration which reflects the non-zero charge density. This effect however
could be neglected since the density of baryon number 
in the D7 brane would be 
negligibly small. 

In the next, we consider the vertex part, which is given by probe D5 brane.}

\subsection{Rotating $D5$ brane vertex}
\label{eqnsec2}

The D5-brane action is thus written by the Dirac-Born-Infeld (DBI) plus
Chern-Simons term
\begin{eqnarray}\label{d5action}
S_{D5}&=&-T_{5}\int d^6\xi
 e^{-\Phi}\sqrt{-\det\left(g_{ab}+2\pi\ap F_{ab}\right)}+T_{5}\int
\left(2\pi\ap F_{(2)}\wedge c_{(4)}\right)_{0\ldots 5}~,\\
g_{ab}&\equiv&\p_a X^{\mu}\p_b X^{\nu}G_{\mu\nu}~, \qquad
C_{a_1\ldots
a_4}\,\equiv\,\p_{a_1}X^{\mu_1}\ldots\p_{a_4}X^{\mu_4}C_{\mu_1\ldots\mu_4}~.\nonumber
\end{eqnarray}
where $T_5=1/(\gs(2\pi)^{5}\ls^{6})$ and $C_{(4)}$ are the brane tension and induced four form respectively.
The Born-Infeld term involves
the induced metric $g$ and the $U(1)$ worldvolume
field strength $F_{(2)}=d A_{(1)}$.
The second term is the Wess-Zumino coupling of the
worldvolume gauge field, and it is also written as
$$
S = -T_5 \int d^6\xi~ e^{-\Phi}
     \sqrt{-\det(g+F)} +T_5 \int A_{(1)}\wedge G_{(5)}~,
$$
in terms of (the pullback of)
the background five-form field strength $G_{(5)}=dC_{(4)}$, which effectively
endows the fivebrane with a $U(1)$ charge proportional to the
${\bf S}^{5}$ solid angle that it spans.

\vspace{.5cm}
Then we give the embedded configuration of the D5 brane in the given 10D
background by minimizing the action. At first, we fix its world volume as
$\xi^{a}=(t,\theta,\theta_2,\ldots,\theta_5)$.
For simplicity we
restrict our attention to the metric (\ref{background-susy}) 
with the ansatz, $\rho=\rho(\theta)$ and $\tilde{\theta}=\omega t$. 
Next we consider the
$SO(5)$ symmetric configurations of the 
form
$\rho(\theta)$, $z(\theta)$, and $A_t(\theta)$ (with all other fields 
set to
zero), where $\theta$ is the polar angle in spherical coordinates.
The action then simplifies to
\be \label{d3action}
S= T_5 \Omega_{4}R^4\int dt\,d\theta \sin^4\theta \{ -
  \sqrt{e^{\Phi}A^2\left({R\over z}\right)^4\left(1-\rho^2\omega^2\right)
     \left(z^2+z^{\prime 2}+A^2\rho^{\prime 2}\right)
   -F_{\theta t}^2}  +4 A_t \},
\ee
where $\Omega_{4}=8\pi^{2}/3$ is the volume of the unit four-sphere.

The gauge field equation of motion following from this action leads
$$
\partial_\theta D = -4 \sin^4\theta,
$$
where the dimensionless displacement 
is defined as the variation of the
action with respect to $E=F_{t\theta}$, namely
$D=\delta \tilde{S}/\delta F_{t\theta}$ and 
$\tilde{S}=S/T_5 \Omega_{4}R^4$. The solution to this equation 
is
\beq\label{d}
D\equiv D(\nu,\theta) = {3\over 2}(\nu\pi-\theta)
  +{3\over 2}\sin\theta\cos\theta+\sin^{3}\theta\cos\theta.
\eeq
Here, the integration constant $\nu$ is expressed as $0\leq\nu=k/N_c\leq 1$,
where $k$ denotes the number of strings emerging from one of the pole of
the ${S}^{5}$, which is wrapped by the D5 brane vertex. 
Next, it is convenient to eliminate the gauge 
field
in favor of $D$ and Legendre transform the original Lagrangian to
obtain an energy
functional of the embedding coordinate only:
\beq\label{uD5}
U_{\rm D5} = {N\over 3\pi^2\alpha'}\int dz~{R^2\over z^2}e^{\Phi/2}A(z)
\sqrt{\left(1-\rho^2\omega^2\right)
  \left(1+z^2\left({\partial\theta\over \partial z}\right)^2 +A^2
\left({\partial\rho\over\partial z}\right)^2\right)}\,
\sqrt{V_{\nu}(\theta)}~.
\eeq
\beq\label{PotentialV}
V_{\nu}(\theta)=D(\nu,\theta)^2+\sin^8\theta\, ,
\eeq
where we used $T_5 \Omega_{4}R^4=N_c/(3\pi^2\alpha')$, and the integral variable
is changed from $\theta$ to $z$. So $\theta$ is a function of $z$ in this case.
Using this expression (\ref{uD5}) and (\ref{PotentialV}), 
we can solve the spinning D5 brane configurations.

\vspace{.3cm}

In the following, we consider two cusps at $\theta=0$
and $\theta=\pi$, where the displacement is finite and actually we have
\beq
 D(\nu,0)={3\over 2}\nu\pi\ , \quad D(\nu,\pi)={3\over 2}(\nu-1)\pi\, .
\eeq
They should be cancelled by the fundamental strings to form a baryon.
We use the above setting of the D5 brane for 
two typical baryon states. {(a) One
is set as $\nu=0$ or 1 and $\rho=0$. 
We call this state as point vertex baryon.
The D5 brane is seen as a point in our real
three space since it does not extend in the direction of $\rho$. (b) Another 
configuration is obtained for $0<\nu <1$ and $\partial \rho / \partial
z\neq 0$.} In this case, the D5
brane is extended as a string in the direction of $\rho$, and the quarks are
attached on both side of this extended vertex. 
The quarks are separated to $N_c\nu$ and $N_c(1-\nu)$ quarks
on the end points of the vertex. This is called as the split vertex baryon.
The configurations of these two types of baryonic states are shown in the Figs.
\ref{d5-d7-point} and \ref{d5-d7-split}. In the next, we examine the spin and the energy of these
configurations.

\vspace{.3cm}
\subsection{Spin and Energy for Baryon}

{\bf (a) Point vertex;}

\begin{figure}[htbp]
\vspace{.3cm}
\begin{center}
\includegraphics[width=12cm]{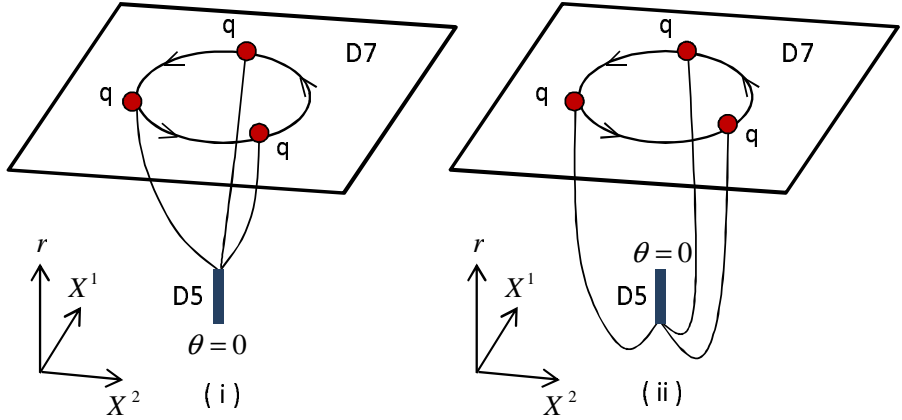}
\caption{{\small Rotating baryon configurations for point vertex.}}
\label{baryon-config2}
\end{center}
\end{figure}

In this case, the D5 brane vertex contributes only to the energy and
not to the spin $J$, which
is obtained from the strings attached to the D5 brane point-vertex.
Then we have
\beq\label{point-B}
  J_P=N_c J_s\, , \quad E_P=N_c E_s +U_{\rm D5}(\rho=0)|_{\nu=0}\, ,
\eeq
where $J_s$, $E_s$ and $U_{\rm D5}$ are given by (\ref{spin}), (\ref{energy})
and (\ref{uD5}). {The configuration of D5 brane for
$\nu=1$ is the same as the one of $\nu=0$ case, so it is sufficient to
consider only the above case.} 
$J_P$ and $E_P$ are calculated as follows. At first, solve
the equation of motion for $z(\theta)$ by using $U_{\rm D5}(\rho=0)|_{\nu=0}$.
Then we find the tension of the D5 brane at $\theta=\pi$, and this is balanced
by the string part, and this is expressed by the ``no-force condition'' \cite{GI}.
This condition provides the boundary condition for the equations of motion
of strings given by (\ref{string-ac0}). 
It is obtained at $\rho=0$ and $\theta=\pi$ as,
\beq\label{no-force}
  \left({\partial z\over \partial \rho}\right)_{\rm string}
   =\left({1\over z}{\partial z\over \partial \theta}\right)_{{}_{\rm D5}}\, .
\eeq
This condition gives the balance of the force in the $z$ direction
at the connecting point of the strings and the D5 brane. The 
condition for the force in the $\rho$ direction is satisfied by putting the
strings symmetrically around the D5 brane vertex as shown in the Fig.\ref{baryon-config2}. 

As shown in \cite{GI}, we find two typical string configurations in the 
case of the point vertex. They are discriminated by the conditions,
(i) $1/z(0)<1/z(\pi)$ or (ii) $1/z(0)>1/z(\pi)$ of the D5 brane configuration.
They are shown in the Fig.\ref{baryon-config2}. 

\vspace{.3cm}
For the case of (i), the strings can not stretch so much, then this configuration
contribute only to the small $J_P$ and small $E_P$ baryons. 
{In this case, the lowest energy of the baryon is obtained when 
the strings shrink up to the D7 brane, and then the baryon is almost
constructed by the D5 vertex. This configuration gives the lowest baryon
mass, which is finite contrary to the lowest zero meson mass. Here we give
the lower bound for this lowest baryon mass. It is obtained from 
the D5 brane which doesn't stretch to $z$-direction ($\partial z/\partial \theta=0$).
This is realized when the position of the D5
brane ${z=z_{{\rm D5}}}$ is located at the same point of the D7 brane $z=z_{{\rm D7}}$, namely $z_{{\rm
D5}}=z_{{\rm D7}}$. 
Such a solution is given \cite{GI} as follows, 
\beq
z_{{\rm D5}}=z_{{\rm D7}}=\tilde{q}^{-\frac{1}{4}}.
\eeq
\begin{figure}[htbp]
\vspace{.3cm}
\begin{center}
\includegraphics[width=12cm]{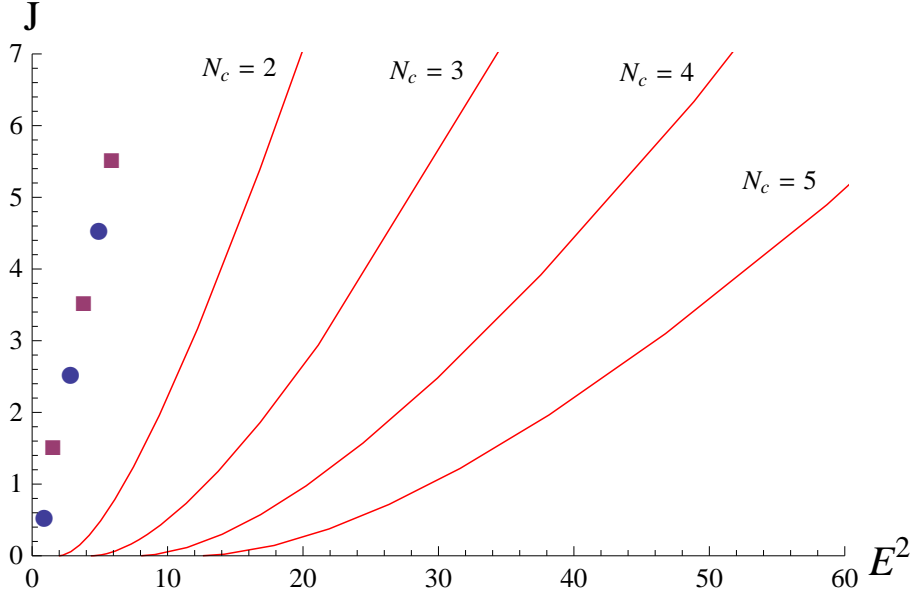}
\caption{{\small The baryon trajectories with point
 vertex are shown for $N_c =2\sim 5$. 
The curves are given for ${\alpha}'_M = 0.88$ (GeV$^{-2}$), $m_q = 0.3$ (GeV), and $\lambda= R^2/{\alpha}'=10$.
The squares and dots denote the $\Delta$ and $N$ baryon series respectively.}}
\label{baryon-point1}
\end{center}
\end{figure}
For the above constant-point solution, we have
\beq
  U_{D5}
  ={\sqrt{2}N_c\over 3\pi^2}{\alpha_M'}^{-1/2}\lambda^{1/4}j(0)\, , \quad
  j(\nu)=\int_0^{\pi}d\theta \sqrt{V_{\nu}(\theta)}\, ,
\eeq
where $j(0)=7.6$. When we use the experimental relation,
$\alpha_M'=0.88$ (GeV$^{-2}$) 
we find
\beq
  U_{D5}=0.387\times N_c\lambda^{1/4}\, .
\eeq
In general, $z_{{\rm D5}}>z_{{\rm D7}}$, then for $\lambda >1$, we have the lowest mass bound, which 
exceeds the nucleon mass for $N_c=3$.}

\vspace{.3cm}
For large $J_P$ and $E_P$, the baryons are expressed by the
the configuration (ii), we can approximate $E_P$ as
\beq
  E_P\simeq N_c E_s \simeq N_c {(1+\beta)\over 2}E_M\ ,
\eeq
since the vertex part $U_{\rm D5}(\rho=0)|_{\nu=0}$ is negligible compared
to the string part which is stretched long. 
The second equality is obtained since one
of the $N_c$ strings can be replaced by the half and some extra ($\beta$) part
of the meson (see (ii) of the Fig.\ref{baryon-config2}). Then we find 
following linear relation between $J_{\rm P}$ and $E_P^2$,
\beq\label{t-P-Baryon}
 J_{\rm P}\simeq {2\over N_c(1+\beta)}{1\over 2\pi\tau_M}{E_P}^2\, .
\eeq

Then the slope of the linear trajectory 
at large $E_P$ is given as
\beq\label{Baryon-slope1}
 \alpha_B^{Regge}={2\over N_c(1+\beta)}\alpha_M^{Regge}\, .
\eeq
This is smaller than $2/N_c\times \alpha_M^{Regge}$, 
which is expected from naive quark model. According to our numerical analysis,
it is about 60 percent of the expected one for $N_c=3$ here. The reason of this
suppression can be reduced to factor $1/(1+\beta)$ in equation (\ref{t-P-Baryon}).
In the present case, we find $\beta\sim 1/8$ in our numerical
calculation given here, but this factor will depends on 
the configurations of the rotating strings.
 
On the other hand, the experimental result given by the nucleon series
implies $\alpha_B^{Regge}\simeq 0.99$ (GeV$^{-2}$). This can not be obtained
from the above result (\ref{Baryon-slope1}). 
Then we should consider another 
baryon configuration for higher spin states.

\vspace{.3cm}

\noindent{\bf (b) Split vertex;}

\vspace{.3cm}
In the next, we consider the split vertex, and the baryon is expressed almost
by the vertex only. In other words, the strings are absorbed into the vertex and 
their end points are attached on the two D7 branes. So the baryon configuration considered here
is described by the D5 brane, see Fig. \ref{baryon-config2}. 
Here the two cusps of D5 brane are considered as the 
strings with zero length, then we impose the boundary condition for the cusps,
\beq
  \partial_z\rho|_{z=z_{\rm D7}}=0\, .
\eeq
From (\ref{uD5}) the spin is obtained as 
\beq
 J_{D5}=-{\partial U\over \partial\omega}
   ={N_c\over 3\pi^2\alpha'}\int dz~{R^2\over z^2}e^{\Phi/2}A\rho^2\omega
\sqrt{1+z^2\left({\partial\theta\over \partial z}\right)^2 +A^2
\left({\partial\rho\over\partial z}\right)^2\over 
1-\rho^2\omega^2}\,
\sqrt{V_{\nu}(\theta)}~.
\eeq
We remember that (\ref{uD5}) originally includes the $d{\tilde{\theta}}/dt=\omega$, so it is 
further Legendre transformed with respect to $U$. Then
we obtain the energy
\beq\label{D5-energy}
  E_{D5}=-{\partial U\over \partial\omega}\omega+U
   ={N_c\over 3\pi^2\alpha'}\int dz~{R^2\over z^2}e^{\Phi/2}A
\sqrt{1+z^2\left({\partial\theta\over \partial z}\right)^2 +A^2
\left({\partial\rho\over\partial z}\right)^2\over 
1-\rho^2\omega^2}\,
\sqrt{V_{\nu}(\theta)}~.
\eeq
\begin{figure}[htbp]
\vspace{.3cm}
\begin{center}
\includegraphics[width=7cm]{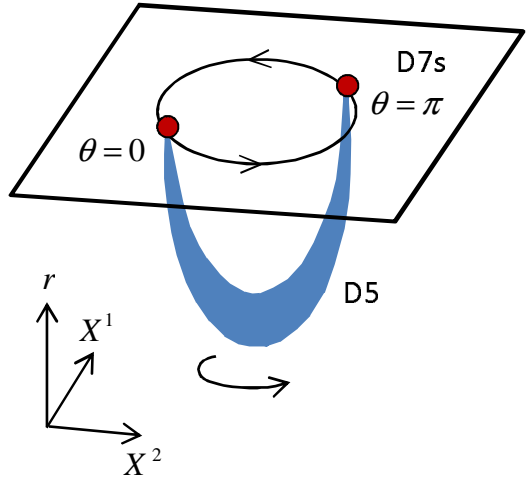}
\caption{{\small A typical rotating baryon configuration for split vertex with
minimum energy is show in the 3D space of ($X^1$,$X^2$,$r$). In this configuration, we can
ignore F-strings which connect the D5 and D7 branes since their
lengths are almost zero. Therefore F-strings do not contribute to the
energy and also to the spin of the baryon. Here 
we notice that the two D7 branes at different $X^9$
are sitting at the same $r$, and the two end points of D5 brane are 
moving on the different
D7 branes. }
}
\label{baryon-config3}
\end{center}
\end{figure}

\vspace{.3cm}
The energy and the spin
are evaluated as in the case of the meson. Firstly, we rewrite
the above equation (\ref{D5-energy}) as
\beq
  E_{D5}=\int_{\rho_L}^{\rho_R} d\rho M_{\rm sta}^{D5}\sqrt{1\over 1-\omega^2\rho^2}\, ,
\eeq
\beq
 M_{\rm sta}^{D5}={N_c\over 3\pi^2\alpha'}e^{\Phi/2}A^2(z){R^2\over z^2}
        \sqrt{1+z^2(\partial_{\rho}\theta)^2+A^{-2}(z)(\partial_{\rho}z)^2}
        \sqrt{V_{\nu}(\theta)}\, ,
\eeq
where $\rho_{L}$ and $\rho_{R}$ denotes the end points of the D5 brane vertex.
For the case of mesons, the end points of the string are given as 
$\rho_{R}=-\rho_{L}=\rho_b$ since the configuration is symmetric. But
in the present case, the D5 brane vertex is not symmetric except for
$\nu=1/2$. In spite of this fact, we find a similar relation of energy and spin.

For the case of large $E_{D5}$, $M_{\rm sta}^{D5}$ can be approximated by the value
at $\theta=\theta_c$, where $V_{\nu}(\theta)$ is maximum. And it is given as 
the solution of the following equation \cite{GI},
\beq
  \pi\nu=\theta_c-{1\over 2}\sin(2\theta_c)\, .
\eeq
Further, we find
\beq
  \partial_{\rho}z=\partial_{\rho}\theta=0\, 
\eeq
at the point of $\theta=\theta_c$. As a result, we obtain
\beq
  M_{\rm sta}^{D5}={2N_c\over 3\pi}\sin^3(\theta_c) \tau_M\equiv \tau_B\, ,
\eeq
where $\tau_M$ is given above Eq. (\ref{t-meson}). Then
\beq
  E_{D5}\simeq\tau_B\int_{\rho_L}^{\rho_R} d\rho \sqrt{1\over 1-\omega^2\rho^2}
             \simeq \tau_B\pi\rho_R\, ,
\eeq
where we used $\omega\rho_R=1$ and $\omega\rho_L=-1$. $J_{D5}$ is similarly
obtained as
\beq
 J_{D5}\simeq {\pi\over 2}\tau_B\rho_R^2={1\over 2\pi\tau_B}E_{D5}^2\, .
\eeq

\vspace{.3cm}
Then we could obtain the similar result to the meson trajectory with the different
slope, 
\beq
 \alpha_B'={1\over 2\pi\tau_B}\, .
\eeq
The experimental data of $\alpha_B'$ lead to a similar value with that of meson, 
$\alpha_B'\simeq \alpha_M'\simeq 1$ (1/GeV$^2$). Then we expect 
\beq
  {2N_c\over 3\pi}\sin^3(\theta_c)\simeq 1\, . 
\eeq
For $N_c=2, 3$, we can not obtain a configuration for the split baryon which
satisfies the no force condition. The reason is that one string must
couple to one of the cusps of D5 brane, and this situation is not
allowed due to the no force condition \cite{GINT2}.
Then,
we find ${2N_c\over 3\pi}\sin^3(\theta_c)=0.85, 1.02$ for $(N_c,\nu)=(4,1/2),
(5,2/5)$ respectively. In these cases, we can set no force condition and satisfactory
slope parameter.
These are assured by the numerical calculation given below with numerical exact
solutions.

\vspace{.3cm}
\noindent{\bf Numerical estimation}

\vspace{.3cm}
In order to obtain the spin and energy, we
solve the equations of motion of the action (\ref{uD5}). It is convenient
to rewrite $U$ in the parametrization invariant form to solve the equations.
\be \label{u2}
U = {N_c\over 3\pi^2\alpha'}\int ds~{R^2\over z^2}e^{\Phi/2}A
\sqrt{\left(1-\rho^2\omega^2\right)
  \left(\dot{z}^2+z^2\dot{\theta}^2 +A^2\dot{\rho}^2\right)}\,
\sqrt{V_{\nu}(\theta)}~.
\ee
where dot denotes the derivative with respect to the introduced parameter $s$, for example $\dot{z}=\partial z/\partial s$. We rewrite $U$ as
\be \label{u3}
U ={N_c\over 3\pi^2\alpha'}\int ds~\tilde{U}\, , 
\ee
where
\beq 
  \tilde{U}=Q\sqrt{\dot{z}^2+z^2\dot{\theta}^2 +A^2\dot{\rho}^2}\, , \quad
   Q={R^2\over z^2}e^{\Phi/2}A
\sqrt{\left(1-\rho^2\omega^2\right)V_{\nu}(\theta)}\, .
\eeq

Then the canonical momenta of $q_i(=(z,\theta,\rho))$ are given by 
$p_{i}={\partial \tilde{U}\over \partial \dot{q_i}}$ as
\beq
 p_{z}=
{\dot{z}\over\Delta}\, , \quad
 p_{\theta}=
z^2{\dot{\theta}\over\Delta}\, , \quad ,
  p_{\rho}=A^2{\dot{\rho}\over\Delta}\, ,
\eeq
where
\beq \label{delta}
\Delta=\frac{\sqrt{\dot{z}^2+z^2\dot{\theta}^2 +A^2\dot{\rho}^2}}{Q}\, .
\eeq
Then the Hamiltonian $H$ is written as
\beq
 H=\sum p_{i}\dot{q}_i-\tilde{U}
=\Delta\left(p_z^2+{p_{\theta}^2\over z^2}+{p_{\rho}^2\over A^2}-Q^2\right)\, .
\eeq
Due to the reparametrization invariance of the system, we have the constraint
$H=0$. This constraint can be changed to a more convenient form by changing
the Hamiltonian as
\beq
 H=2\tilde{H}\Delta\, , \quad
 \tilde{H}={1\over 2}\left(p_z^2+{p_{\theta}^2\over z^2}+{p_{\rho}^2\over A^2}-Q^2\right)\, ,
\eeq
and the equations of motion are obtained for new Hamiltonian $\tilde{H}$ as
\beq\label{dot-r-th}
\dot{z}=p_{z}\, , \quad
\dot{\theta}={p_{\theta}\over z^2}\, , \quad 
\dot{\rho}={p_{\rho}\over A^2}\, ,
\eeq
\beq\label{dot-pr-th}
 \dot{p}_z=-{\partial \tilde{H}\over \partial z}\, , \quad
 \dot{p}_{\theta}=-{\partial \tilde{H}\over \partial \theta}\, , \quad
 \dot{p}_{\rho}=-{\partial \tilde{H}\over \partial \rho}\, .
\eeq

\vspace{.3cm}
\begin{figure}[htbp]
\vspace{.3cm}
\begin{center}
\includegraphics[width=7cm]{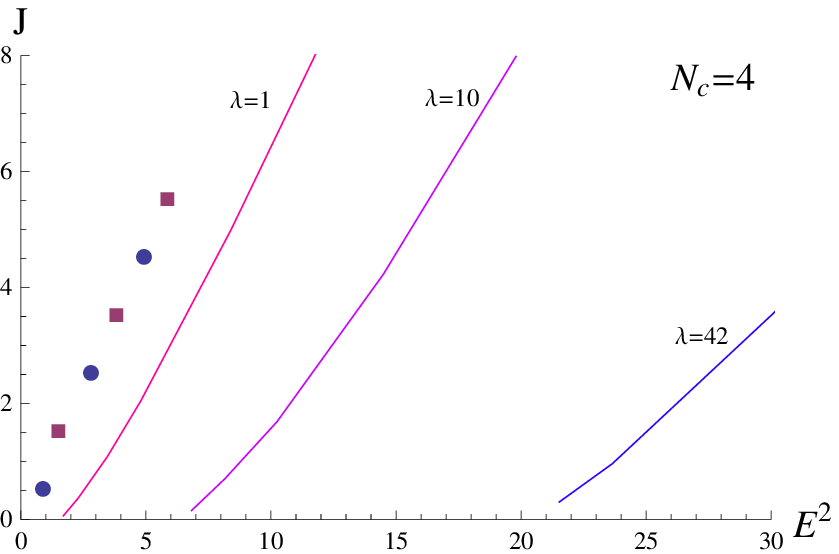}
\includegraphics[width=7cm]{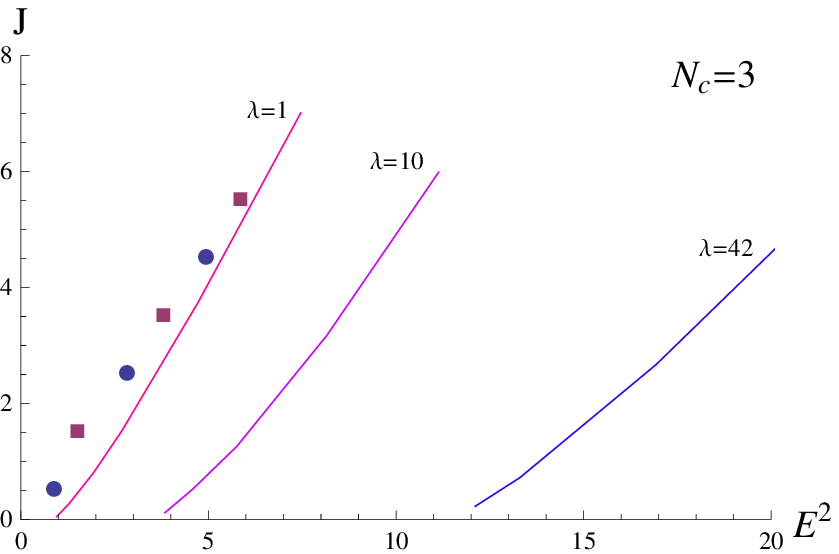}
\caption{{\small The baryon trajectories with rotating D5 split
 vertex are shown for $N_c =4$ (left) and $N_c =3$ (right). 
The curves are given for ${\alpha}'_M = 0.88$ (GeV$^{-2}$), $1/z_{D7} = 
 1.0$ (GeV), and three values of $\lambda= R^2/{\alpha}'$.
The large and small dots denote the $\Delta$ and $N$ baryon series
 respectively.}}
\label{baryon-D5rot}
\end{center}
\end{figure}

{In Fig. \ref{baryon-D5rot}, we show our numerical results of the split 
baryon trajectories for the cases of
$\left\{N_c=4,~ \nu=1/2\right\}$ and $\left\{N_c=3, ~\nu=1/3\right\}$.
In each case, the calculations are given for $m_q(=\sqrt{\lambda}/(2\pi z_{D7}))=0.16,~0.51,~1.04$ (GeV). }
The slope parameter
is compatible with the experimental spectra. However, 
in the case of baryon, the masses become very large for large $\lambda$ due to the vertex energy. 
The vertex does not disappear when the distance between two ends of split vertex
becomes zero, and then it remains to give the lowest mass
of the baryon. While we expect that this lowest mass would be near the nucleon mass,
it depends on $\lambda$ as shown in the Fig. \ref{baryon-D5rot}. Since
the holographic approach is trusted for $\lambda \gg 1$, it seems to be difficult 
to get a lowest mass comparable to the nucleon mass. 
{It may be
about three or four times of the nucleon mass for $\lambda\sim 5 >1$. }
We should notice that the meson trajectory 
is not changed by the parameter $\lambda$ as shown above.

Another problem is that we can not get the no-force condition for the split baryon
of $N_c=3$ \cite{GINT2}. 
The three F-strings going out from the two pole-points of split vertex
are separated to one and two in the case of $N_c=3$. We impose the no-force conditions
on both sides where the F-string(s) and D5 brane couple. However, we know that on the
side of one F-string it is impossible to balance the force due to the tension of the
one F-string and the one of the D5 brane since the former one is always greater than
the latter. As a result, the string shrinks up to zero length and disappears. Then this 
cusp of the D5 vertex can not be in a stable state. So we should extend to $N_c=4$
to have a stable split vertex. 
{This is the reason why we show the example of
numerical simulation only for $N_c=4$ at the same time.}

{One way to allow the baryon state of $N_c=3$ and $\nu=1/3$ is to 
introduce a pair of quark and anti-quark at the cusp with one F-string coupled.
Then the no force conditions are satisfied in this case, but there appears
another possible instability to annihilate the introduced a pair of quark and anti-quark. The energy to make this vertex state is however very small since
the cusp is nearly at the flavor brane position, so we can consider this state
within an energy fluctuation of quantum corrections which would be important
at this cup point. In this sense, it is meaningful to consider the split vertex
also for $N_c=3$

As a result, we could say that the higher spin baryon states are realized 
by rotating vertex, which extends as a string with a definite tension.
The F-strings, which are attached to the vertex, shrink to almost zero length
and don't contribute to the spin of the baryon state. This picture of baryon
configuration is different from the one considered by the naive quark model.
We could find that the main ingredient of the baryon is not the F-string but
the vertex, see Fig. \ref{baryon-config3}.}

\section{Summary and Discussion}

Using a holographic model for mesons and baryons with higher spins, Regge behaviors are studied.
In our model, confinement is realized due to the gauge condensate 
$\langle F^2\rangle$ which determines the tension $\tau_M$ of the linear potential
between the quark and anti-quark as $\tau_M=\sqrt{\langle F^2\rangle}/2$.
As a result, the slope of the meson trajectory is given by this quantity 
as $\alpha'_M=1/(\pi\sqrt{\langle F^2\rangle})$, and then $\langle F^2\rangle$
is fixed from the experimental data of meson spectrum. We obtain 
 $\langle F^2\rangle=0.13$ (GeV$^4$) from the experimental data of
$\rho$ meson trajectory. On the other hand,
the intercept of the trajectory is always zero in our model since the meson
is approximated by a classical configuration of Nambu-Goto string, which is embedded
in a bulk dual to the confining gauge theory. In this model, the string
is reduced to a point when its energy approaches to zero. Then the angular 
momentum also tends to zero. This would be improved by considering
the quantum effects, which are not considered here.

As for the baryons, the Regge behavior is examined for two typical configurations,
which are discriminated by the vertex. The baryon vertex is given by D5 brane which is introduced in the bulk background as a probe brane with $U(1)$ flux. It wraps on $S^5$ in the bulk. We consider two simple vertex configurations. One 
is given by a solution being extended to the fifth coordinate $r$ as $r(\theta)$.
In this case, the vertex is observed as a point in our 4D space-time, so we call this as a point vertex. Another configuration is 
obtained as a solution of the equations of motion of 
two coordinates, $r(\theta)$ and $x(\theta)$. This configuration
extends like a string in the plane of $r-x$, where $x$ is one of our three space coordinates. This is called as split vertex since the vertex is observed as a line extended in the $x$ direction. (See also Figs.\ref{d5-d7-point} and
\ref{d5-d7-split}) 
{We should notice that we need at least two flavor D7 branes
for the split vertex baryon as shown in 
the Fig. \ref{d5-d7-split}. This implies that the number of the flavor quarks
might be related to the number of the cusps on the D5 brane
baryon vertex. We should examine this point
more in the future.}

In the case of the point vertex (see Fig. \ref{baryon-config2}), spin and energy
are mainly given by the strings. Then the slope of the baryon,
$\alpha'_B$, is  related to the 
meson slope as in the form $\alpha'_B=2\alpha'_M/(N_c(1+\beta))$, where 
$\beta$ denotes a positive factor which depends on the string configuration. Then, in general,
$\alpha'_B<\alpha'_M$ for $N_c\geq 3$. As a result, we could not obtain 
the slope $\alpha'_B$, which is comparable with the experimental data.

On the other hand, for the split vertex case,
the spin and the energy are supposed to be
provided by the vertex only since the energy is minimized 
for this configuration (see Fig. \ref{baryon-config3}). Then we can neglect the contribution from the strings
in the analysis.
The vertex with higher spin, in this case, extends
in one spatial direction $x$ of our three dimensional space. 
Then the situation is similar to the case of mesons. That is to say that
the slope parameter is determined by the tension of this string-like vertex when
it extends in the $x$ direction. This tension
is in general smaller than the summation of the independent
$N_c$ strings, then the slope of the
trajectory becomes larger than that of the point-vertex case. Actually we 
could find a slope,
which is compatible with the experimental data, for $N_c=3$.
In this case, the problem is that
we must adjust the parameters to obtain the lowest
mass of the baryon compatible with the nucleon mass. This is performed 
by using
unfavorably small $\lambda$.
Furthermore, 
in order to satisfy the no-force condition at the vertex, we should extend 
$N_c$ to a large value $N_c\geq 4$. However, this extension moves the baryon mass to the large side. These points should be resolved by including quantum
corrections. 
{Another possible improvement will be found in other holographic model
of different bulk background and probe branes.}
They are remained as future works.

\vspace{.3cm}
\section*{Acknowledgments}
{The authors thank to Akihiro Nakamura 
for useful discussions at the early
stage of this work. F. T thanks to H. Yoneyama and K. Kanaya for a discussion
on the lattice simulations.
T. T. would like to thank to Kouki Kubo and Naoki Yamatsu for helpful comments.}



\newpage
\end{document}